\documentclass{rmf-d}
\usepackage{nopageno,rmfbib,multicol,times,epsf,amsmath,amssymb,cite}
\usepackage[]{caption2}
\usepackage{color}
\usepackage[dvipsnames]{xcolor}

\usepackage{appendix}
\usepackage{soul}
\usepackage{ulem}
\usepackage{graphicx}
\usepackage{graphics}
\usepackage{tikz}
\usepackage{longtable}
\usepackage{comment}
\usepackage{hyperref}
\def\rmfcornisa{Research OR Education of Physics \hfill\rmf\ {\bf ??} (*?*) ???--???
\hfill MES? A\~NO?}

\def\rmfcintilla{{\it Rev.\ Mex.\ Fis.\/} {\bf ??} (*?*) (????) ???--???}
\clearpage \rmfcaptionstyle 
\begin{document}
\title{   Meson and Glueball spectroscopy within the Graviton Soft-Wall model
\vspace{-6pt}}
\author{ Matteo Rinaldi     }
\address{ Dipartimento di Fisica e Geologia. Universit\`a degli studi di 
Perugia. INFN section of Perugia. Via A. Pascoli, Perugia, Italy.    }
\author{ }
\address{ }
\author{ }
\address{ }
\author{ }
\address{ }
\author{ }
\address{ }
\author{ }
\address{ }
\maketitle
\recibido{day month year}{day month year
\vspace{-12pt}}
\begin{abstract}
\vspace{1em} In this contribution we present  
results of the calculations of several hadronic spectra within the holographic
graviton soft-wall (GSW) model. In particular, we studied and compared with data
for  the ground state and 
 excitations of:  glueballs, scalar, vector, axial and  pseudo-scalar 
mesons. The GSW model is found to be capable to describe these observable
with only few parameters.

  \vspace{1em}
\end{abstract}
\keys{  PLEASE PROVIDE SUMMARY IN ENGLISH  \vspace{-4pt}}
\pacs{   \bf{\textit{PLEASE PROVIDE }}    \vspace{-4pt}}
\begin{multicols}{2}

\section{Introduction}

In this contribution we investigate
the spectra of some hadronic species. 
We consider the graviton soft-wall (GSW) model, introduced and applied in 
Refs. \cite{Rinaldi:2017wdn,Rinaldi:2021dxh,Rinaldi:2018yhf,Rinaldi:2020ssz}, 
initially adopted
 to describe 
non perturbative features of 
glueballs. We remind that 
the holographic approach relies in a correspondence between a five
 dimensional 
classical theory with an AdS metric and a supersymmetric conformal 
quantum 
field theory. Since the latter is not 
QCD, we use the so-called ``bottom-up'' approach
~\cite{Brodsky:2003px,DaRold:2005mxj,Karch:2006pv}. 
where the five dimensional 
classical theory is properly modified to reproduce non-perturbative  QCD
properties  as 
much as possible. Furthermore, the GSW model is 
 a modification of the initial
soft-wall (SW)  where  a dilaton field  is introduced to softly 
break 
conformal invariance. In GSW model, in order to properly describe the scalar 
glueball spectrum, a modification of the metric has been proposed.
This model has been successfully applied to reproduce non 
perturbative features of mesons and 
glueballs
\cite{Rinaldi:2017wdn,
Karch:2006pv,Capossoli:2015ywa,Erlich:2005qh,Colangelo:2008us}. 
In particular, we 
calculated and impressively described the spectra of: glueballs (with even and 
odd spin), the light and heavy scalar mesons,
the  $\rho$, the $a_1$, the $\eta$  and the $\pi$.
Moreover, we showed in Ref. \cite{Rinaldi:2018yhf}  that
only when the masses of the glueballs and the mesons 
are close, mixing is to be expected~\cite{Vento:2004xx}. However, 
if this mass condition is
 associated  to a  different dynamics, mixing 
will not happen~\cite{Vento:2015yja}.

\section{Essential features of the GSW model}
\label{sec2}
The essential 
difference between the GSW model from the traditional SW one, is a
deformation of the $AdS$ metric in 5 dimensions:

\vskip -0.6cm
\begin{align}
ds^2= e^{\alpha {\phi_0(z)}} g_{M N} dx^M 
dx^N   = \dfrac{R^2}{z^2}\big(\eta_{\mu \nu} dx^\mu dx^\nu-dz^2  \big)
\label{metric5}
\end{align}
where $g_{MN}$ is the AdS$_5$ metric
and $\phi_0(z)= k^2 z^2$
\cite{Erlich:2005qh,Colangelo:2008us,deTeramond:2005su,Rinaldi:2017wdn,
Colangelo:2007pt,Capossoli:2015ywa}.
Modifications of the metric have been also proposed in other studies of the 
properties of mesons 
and glueballs within AdS/QCD 
~\cite{Colangelo:2007pt,Capossoli:2015ywa,Vega:2016gip,Akutagawa:2020yeo,
Gutsche:2019blp, Klebanov:2004ya,MartinContreras:2019kah,FolcoCapossoli:2019imm,
Bernardini:2016qit,Li:2013oda}. 
 The action, in the gravity sector, written in terms 
of the standard AdS metric 
of the  SW model,
is:

\vskip -0.5cm
\begin{align}
\label{prefactor}
 \bar S = \int d^5xe^{\phi_0(z) 
\left(\frac{5}{2} \alpha -\beta+1  
\right) } e^{-\phi_n(z)}\sqrt{-g} e^{-\phi_0(z)} \mathcal{L}(x_\mu,z)~.
\end{align}
 In the GSW model, the parameter $\alpha$ encodes the effects due to the 
modification of the metric, while, $\beta$
  is used to recover the SW results as much as possible. Indeed,
$\beta$ is not a free parameter and it is fixed to lead the SW
kinematic term in the 
action~\cite{Rinaldi:2017wdn,Rinaldi:2018yhf,Rinaldi:2020ssz}.
For example, for scalar fields $\beta= \beta_s= 1+\frac{3}{2} \alpha$
and for a vector $\beta=\beta_v=
1+\frac{1}{2}\alpha$. In Ref. \cite{Rinaldi:2021dxh}, an
 additional dilaton $\phi_n$ has been also phenomenologically proposed to 
describe scalar and pseudo-scalar mesons in order to obtaina binding potential 
in the equation of motion. We anticipate that this quantity  does not
contain any free parameter,
which are namely $\alpha$ and $k$.   In closing, in order to properly take into 
account the  chiral symmetry breaking, essential to describe the pion spectrum, 
{a modification of the dilaton has been proposed.   }

\section{The glueball spectra within the GSW model  }
\label{secglue}

{ In this section  we discuss and present
the GSW predictions for the glueball spectra with even and odd spins together
with the
 successful comparison 
with lattice
data.}

\subsection{Scalar glueballs }
Here we recall the main results discussed in Ref.\cite{Rinaldi:2017wdn}. 
We remind that in this case, 
 the GSW model predicts that  the
scalar glueball is described by its dual  graviton
 which is a solution of  the Einstein equation (Ee) for a perturbation the
metric
(\ref{metric5}).  The linearized Ee can be rearranged in a Schr\"odinger 
like 
equation:

\vskip -0.5cm
\begin{equation}
-\frac{d^2 \phi(t)}{d t^2} +
 \left(\frac{8}{t^2} e^{2 t^2} - 15 t^2 + 14 - 
\frac{17}{4 t^2}  \right) \phi(t) = \Lambda^2 \phi(t).
\label{Gexact}
\end{equation}
 where, as usual, we assumed  factorization between the $z$ and $x_\mu$ 
dependence, $ 
t=\sqrt{\alpha k^2/2}\; z$ and 
$\Lambda^2 = (2 /\alpha k^2)\; M^2$, being $M$ the mode mass.

It is remarkable that
the potential  is uniquely determined by the 
metric and its modification. The only  free parameter is the scale
factor depending on $\alpha k^2$. This term  is fixed from the comparison with
 lattice
QCD \cite{Rinaldi:2017wdn}.
As one can see in the left panel of Fig. 
\ref{sgsm}, for  $\alpha k^2\sim (0.37$ GeV)$^2$ the  linear
glueball spectrum is well reproduced, at variance with the SW model.
We also stress the good agreement with
the graund state mass obtained by the BESIII data of the $J/\Psi$ decays
~\cite{Sarantsev:2021ein,Klempt:2021nuf} (SDTK) very recently, after analysis.

\subsection{Spin dependent glueball spectra}
\label{sec5}
We found out that the ground state of glubealls with spin is well reproduced
if we 
consider the approach  of Refs.
\cite{Capossoli:2015ywa,BoschiFilho:2005yh,FolcoCapossoli:2019imm} to describe 
spin effects.
In this case the action is
that of a scalar field \cite{Rinaldi:2020ssz}:

\vskip -0.6cm
\begin{align}
\label{scal}
 \bar S &= \int d^5x~\sqrt{-g}e^{-k^2z^2} 
\Big[ g^{MN}\partial_M G(x) \partial_N 
G(x)+ 
\\
\nonumber
&+
e^{\alpha k^2 z^2} M_5^2 R^2 G(x) \Big]~,
\end{align}
and the spin dependence is encoded in the 5-dimensional mass term: $i)$
$ M_5^2R^2 = J(J+4)$ for even spin $J$
and $ii)$
$
M_5^2 R^2 = (J+2)(J+6)$ {for odd spin}~$J$. One should notice that
since $M_5^2R^2 \geq 0$,
the potential in equation of motion is binding and therefore no additional 
dilatons are needed. Results  of the
calculations for the odd 
and even glueballs are shown in Tables I-II, respectively.
As one can see, results are in fairly 
 agreement with data.
We also evaluated and compared
the Regge trajectories provided by the GSW model with lattice data.
In general,
the  form is $J \sim a_g M^2+b_g $, where $g=o$ stands for odd spin and $g=e$
is referred to even spin, respectively. In the odd case: $a_o=0.18 \pm
0.01,~b_o=-0.75 \pm0.28$  \cite{Rinaldi:2021dxh}, in agreement
with $J \sim 0.18M^2+0.25$
\cite{LlanesEstrada:2005jf}. For even glueballs: $a_e =0.21 \pm
0.01,~b_e=0.58 \pm 0.34$  \cite{Rinaldi:2021dxh}
in agreement with $J \sim 0.25 M^2$\cite{Landshoff:2001pp,Meyer:2004jc}.

\section{Meson spectroscopy}
Here we show the main results for the spectroscopy of the $f_0$, heavy scalar,
 $\rho$,
$a_1,~\eta$ and pion mesons.
We stress again that for the latter case a modification of the dilaton $\phi_0$
must be included to incorporate chiral symmetry breaking.

\subsection{Light and heavy scalar mesons}
\begin{figure*}[h]
\includegraphics[scale=0.53]{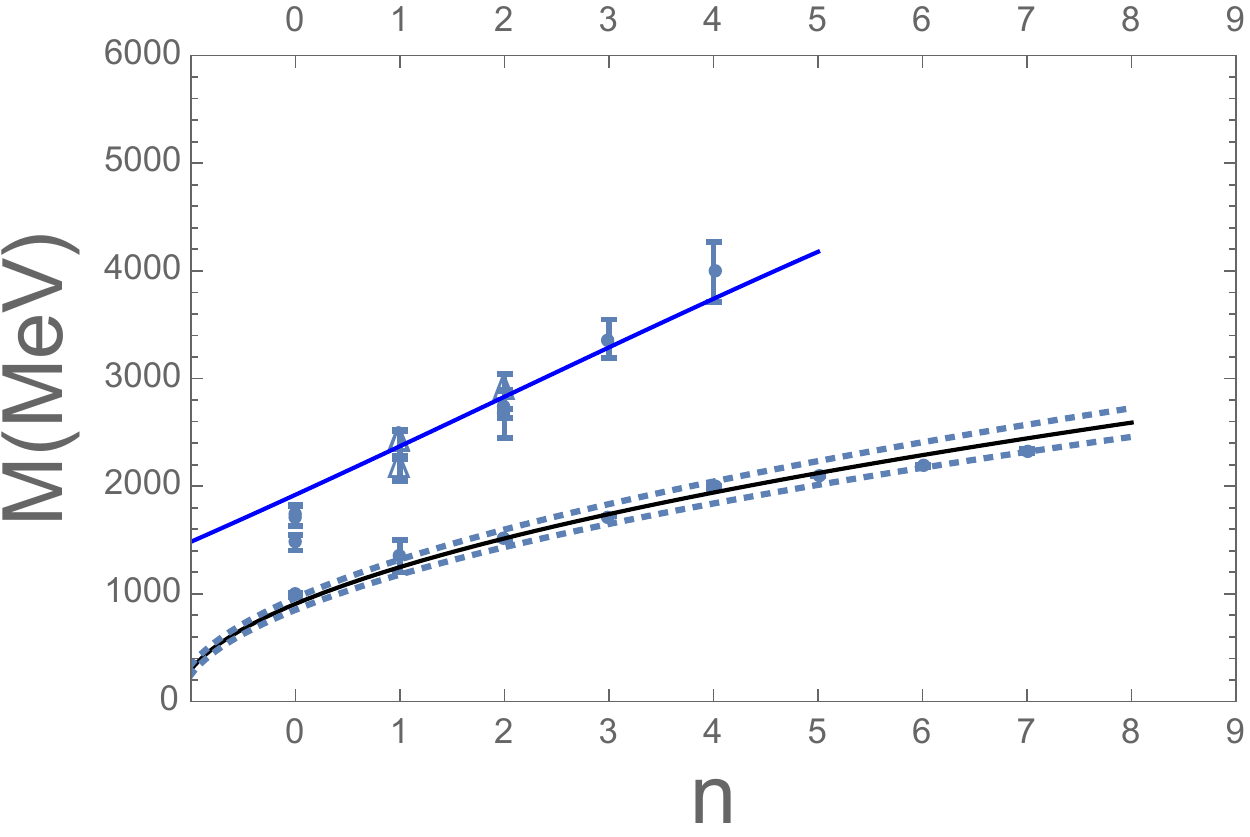} \hskip 1.2cm
\includegraphics[scale=0.55]{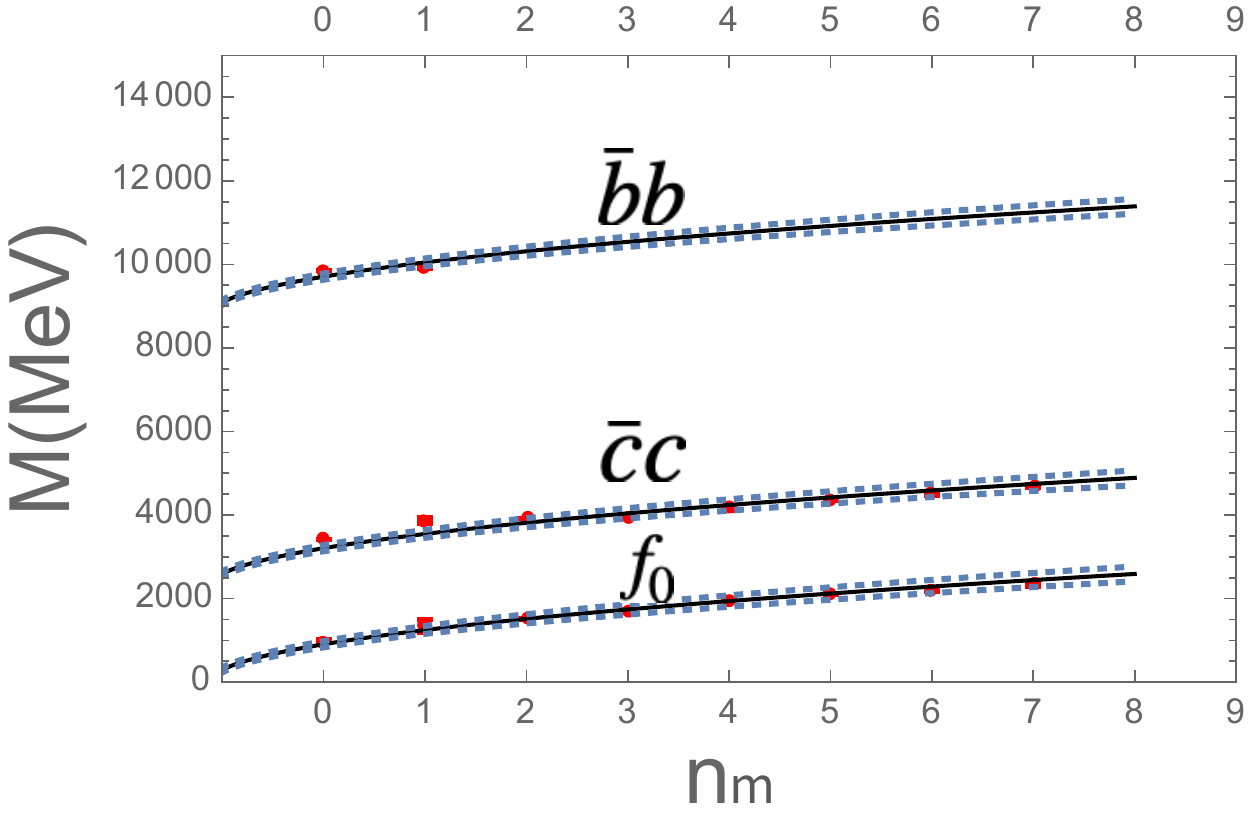}
\caption{\footnotesize  \textsl{
Left panel: GSW fit to the scalar lattice glueball 
spectrum~\cite{Morningstar:1999rf,Chen:2005mg,Lucini:2004my} and to the
experimental scalar meson spectrum~\cite{Tanabashi:2018oca,Zyla:2020zbs}. Solid
line for
$\alpha$: $0.55$  and $0.55\pm0.04$
(dotted). Right panel: The scalar meson spectrum GSW fit to the data  shown for 
all quark sectors. Experimental data
~\cite{Tanabashi:2018oca,Zyla:2020zbs}. The  curves correspond to 
 $C_c = 2400$ MeV for the $c \bar{c}$ mesons
and $C_b = 8700$ MeV for the $b \bar{b}$ mesons. }}
\label{sgsm}
\end{figure*}

In this case the action is that of  Eq. (\ref{scal}) but
since now $M_5^2R^2 = -3$, the relative potential is not binding. Therefore,
at variance to the the glubeall case, the additional contribution
$\exp[-\phi_n(z)]$ must be included.
Details on this topic are presented in Ref. \cite{Rinaldi:2021dxh}. Here we 
mention that $\phi_n$ is chosen to produce in the potential a term proportional 
to the expansion, up to the second order, of $exp(\alpha k^2 z^2)$ and thus 
preserving the binding feature. 
 By keeping fixed $\alpha k^2 =0.37$
GeV$^2$,
 we found a reasonable good fit, see  left
panel of Fig. \ref{sgsm},  for $0.51 \leq \alpha \leq
0.59$. In the case of heavy mesons
we added the quark mass contribution to the light scalar masses
\cite{Rinaldi:2021dxh,Rinaldi:2020ssz} in order to effectively include  the
dynamics the mass of the heavy quarks
\cite{Branz:2010ub,Kim:2007rt,Afonin:2013npa}. In particular, 
the heavy mass ($M_h$) is obtained from the previous light scalar
mass ($M_l$) as follows:
$M_h = M_h + C$,
where $C$ is the contribution of the quark masses.
$C_c=2400$ MeV, for the $ c \bar{c}$ mesons, and for the $ b \bar{b} $
mesons $C_b=8700$ MeV. 
The successful comparison with data  \cite{Rinaldi:2020ssz} is displayed in the 
left and right panels of
Fig. \ref{sgsm}. 
One should notice that
$C_c$ and $C_b$ are comparable with the values of $2
m_c$ and $2 m_b$, respectively, as expected.

\subsection{The $\rho$  spectrum}
\begin{figure*}
\includegraphics[scale=0.65]{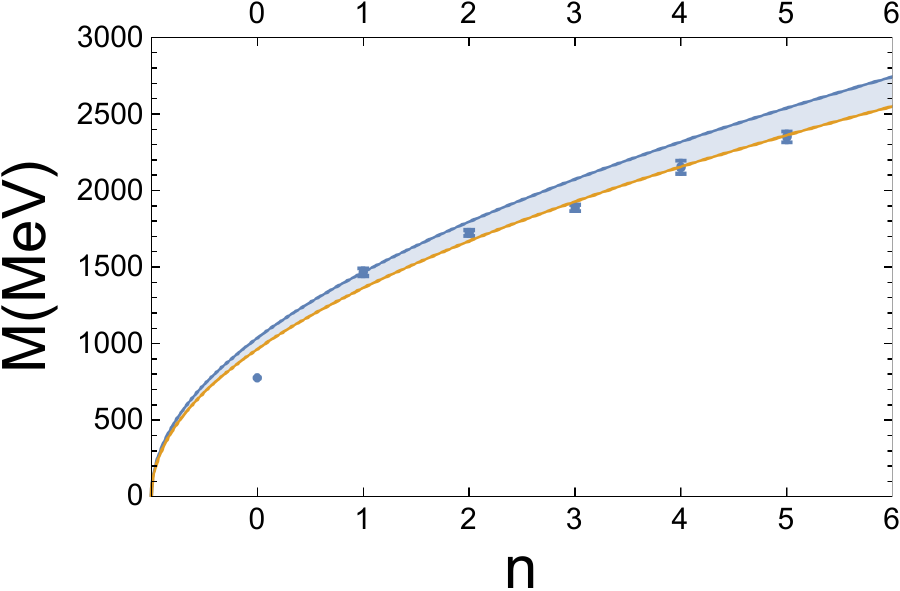}  \hskip 0.2cm
\includegraphics[scale=0.45]{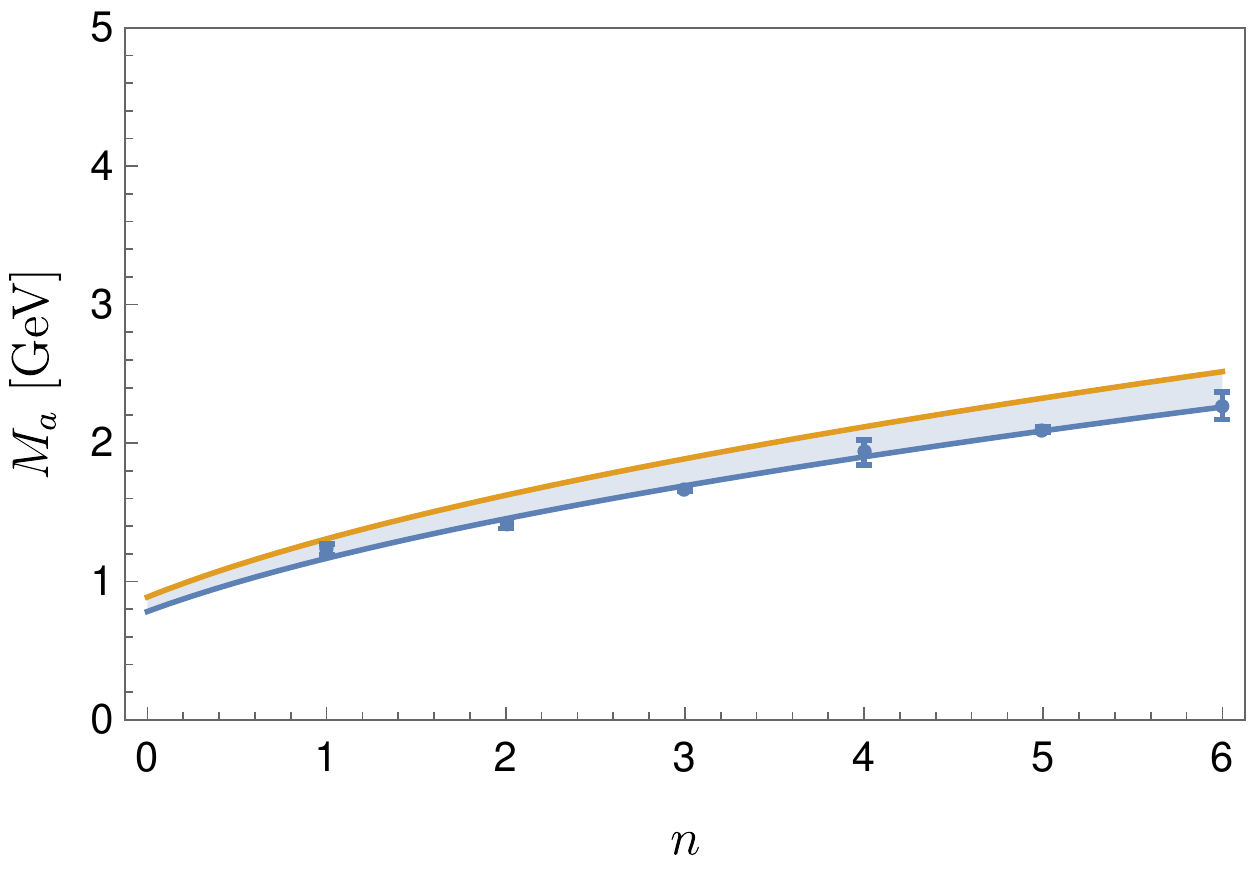}  \hskip 0.2cm
\includegraphics[scale=0.65]{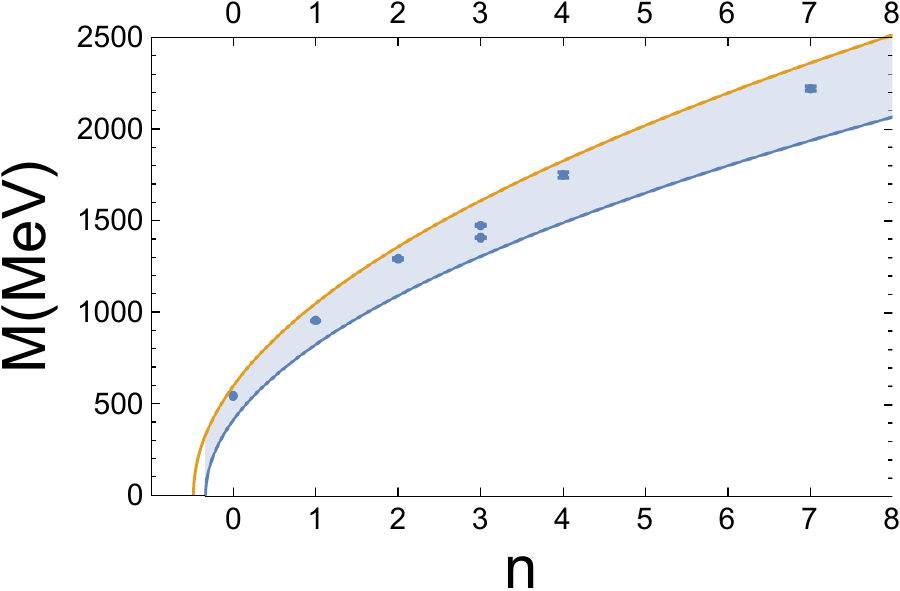}
\caption{Left: The $\rho$ mass plot as a function of mode number.
Experimental data ~\cite{Tanabashi:2018oca,Zyla:2020zbs}.
Center panel, the $a_1$ spectrum. Data from Refs.
\cite{Tanabashi:2018oca,Zyla:2020zbs,Anisovich:2001pn}.
Right panel, the $\eta$ spectrum. Data from Refs.
\cite{Tanabashi:2018oca,Zyla:2020zbs}
}
\label{rho}
\end{figure*}

For the $\rho$ meson, the action is equal to that of a vector field within the
usual  SW model since $M_5^2 R^2=0$ \cite{FolcoCapossoli:2019imm}:

\vskip -0.3cm
\begin{align}
 \bar S = - \dfrac{1}{2} \int d^5x~ \sqrt{-g}~ e^{ -k^2 z^2 } 
\left[ \dfrac{1}{2}  g^{MP}  g^{QN} F_{MN} F_{PQ}  \right]~.
\end{align}
\vskip -0.2
In this case,
there is no need of the auxiliary dilaton since $M_5^2 R^2=0$ and thus the
potential is binding.
As one can see in the left panel of Fig. \ref{rho}, the agreement is good,
exception is $\rho(770)$. Such a discrepancy suggests that the GSW model must
be further improved.

\subsection{The $a_1$ axial meson spectrum}
\label{sec4}

In the case of the  axial-vector mesons, due to
chiral symmetry breaking,  $M_5^2R^2 =-1$ \cite{Huang:2007fv,Contreras:2018hbi}.
 Therefore the EoM for the $a_1$ can be obtained from the action of a vector
field with a conformal mass different from zero:

\vskip -0.5cm
\begin{align}
 \bar S &= -\frac{1}{2} \int d^5x \sqrt{-g} e^{-k^2z^2-\phi_n} 
\left[\frac{1}{2} g^{MP} g^{QN} F_{MN}F^{PQ} \right.
\\
\nonumber
 &+  \left.
 M_5^2 R^2g^{PM} A_P A_M e^{\alpha
k^2 z^2} \right]~.
\end{align} 
\vskip -0.2cm

Since in this case $M_5^2 R^2 < 0$,
the corresponding  potential is not binding and
 a modification of the dilaton is
required. Details on the differential equation defining the contribution
are included in Ref. \cite{Rinaldi:2021dxh}.
With the parameters previously addressed,
 we get the
spectrum shown in the central panel of Fig.\ref{rho}.
Our calculation favors that
the $ a_1(1930),
a_1(2095)$ and $a_1(2270)$ are axial resonances
\cite{Anisovich:2001pn}.
{Moreover,  the agreement is even more impressive if
a missing ground state  with a mass lower
then the
quoted 1230 MeV will be observed.  }

\subsection{The $\eta$ pseudo-scalar meson}
The EoM is similar to that of the scalar case but now,
 $M_5^2 R^2 = -4$ \cite{Contreras:2018hbi}.
As expected,
{one needs to include the additional dilaton  to get a binding 
potential. }
In the right panel of Fig. \ref{rho} we show our calculation of the spectrum, 
we remind that the
band  stands for the
theoretical error on $\alpha$
$\alpha= 0.55\pm 0.04$.  The comparison with  the experimental data
 ~\cite{Tanabashi:2018oca,Zyla:2020zbs} is very good also in this case.
Moreover, the GSW model  predicts that $\eta(1405)$ and
$\eta(1475)$ are degenerate, as
 discussed in PDG review. Moreover, since in the upper mass sector the 
experimental
mass gap is
larger,  the GSW model also favors: {\it i)} the existence of two 
resonances between the $\eta (1760)$ 
and   the 
$\eta(2225)$ and $ii)$ that the $\eta(1405)$ and $\eta(1470)$ are the 
same 
resonance. We recall that the results for the $\rho,~a_1$ and $\eta$ masses are 
free parameter calculations.

\subsection{The pion spectrum}
In the case of the pion,  the Goldstone boson of SU(2)
x SU(2) chiral symmetry,  as already anticipate, we need to incorporate in the 
model  chiral symmetry breaking mechanism.
Since, as 
discussed in e.g., Refs. \cite{Erlich:2005qh,Gherghetta:2009ac,Vega:2016gip}, 
 the physics of confinement and chiral symmetry breaking 
could ascribed to 
 the dilaton ~\cite{Erlich:2005qh,Gherghetta:2009ac,Vega:2016gip}, 
we propose
 a modification of the dilaton profile
function  \cite{Rinaldi:2021dxh}:
 we consider:
\vskip -0.3cm
\begin{align}
\label{49}
\phi_0(z) = \beta_s \tanh{(\gamma z^4 +\delta)} k^2 z^2
\end{align}

\vskip -0.2cm
{The parameter}
$ \tanh{(\delta)}$ is responsible for the
 the chiral symmetry breaking, see details in Ref. \cite{Rinaldi:2021dxh}.
This choice  preserves the
large $z$ behaviour, which leads to
Regge trajectory of the higher mode spectrum.
On the other hand,  the low $z$ region 
describes  the transition region and  $\delta$ and $\gamma$
incorporates the effects of 
{the}  spontaneous chiral symmetry breaking. Also in this case, in analogy with 
the $\eta$,
 we need to include $\phi_n$ to get  a binding potential.
{The  pion spectrum is shown} in the Table III
compared with the PDG data~\cite{Tanabashi:2018oca,Zyla:2020zbs}. 
{The predicts more
pion states  the experimentally observed  
\cite{Contreras:2018hbi,Gherghetta:2009ac}. 
  }

\section{Glueball-Meson mixing}
Here
we discuss the conditions for a not favorable mixing, i.e.
 states with mostly gluonic valence structure \cite{Rinaldi:2018yhf}.
We consider
an
holographic light-fron (LF)
representation of the EoM in term of the
Hamiltonian~\cite{Brodsky:2003px}

\vskip -0.3cm
\begin{equation}
H_{LC} |\Psi_k> = M^2 |\Psi_k>
\end{equation}
and a two dimensional Hilbert space generated by
a meson
and a glueball states, \{$|\Psi_m>, |\Phi_g>$\}.
Mixing occurs when the hamiltonian is not
diagonal in the subspace. A matrix representation of the hamiltonian
is given by

\vskip -0.3cm
\begin{equation}
[H]=  \left( \begin{array}{cc}
m_1 &  \alpha  \\
\alpha & m_2 \end{array} \right) ,
\label{mixing}
\end{equation}
where $\alpha = <\Psi_m|H|\Phi_g>$, $m_1 = <\Psi_m|H|\Psi_m>$
and $m_2 = <\Phi_g|H|\Phi_g>$. We are assuming $m_2>m_1$ and for
simplicity $\alpha$ real and positive.
After diagonalization the eigenstates have a mass
$M_{\pm}= m  \pm \sqrt{ \alpha^2 +
(\Delta m)^2}$,
where $m=(m_1+m_2)/2$ and $\Delta m = (m_2-m_1)/2$.
The first physical meson, assuming to be the
lightest one, is given by the eigenvector of $H$
\cite{Rinaldi:2018yhf}.
Since we fixed the meson spectrum to the experimental values,
 $|\Psi_{phy}>$ represents a physical meson state while we
have fixed the glueball spectrum to the lattice values, therefore the
glueball state
is our initial state $ |\Phi_g>$, thus

\begin{figure*}[htb]

\includegraphics[scale= 0.7]{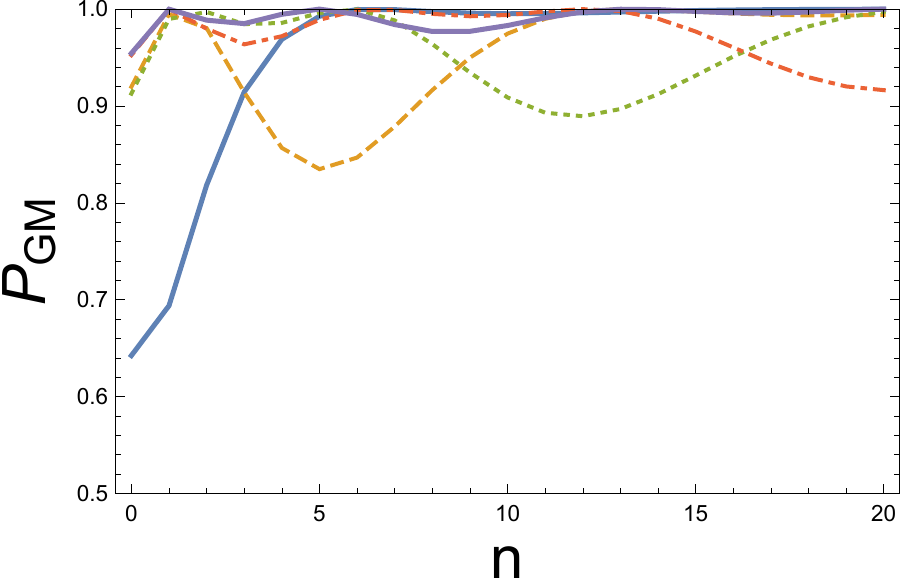}
\vskip -0.2cm
\caption{ \footnotesize{  The probability of no
 mixing for the glueball with
mode numbers
 $n_g = 0$ (solid),$1$ (dashed), $2$ (dotted), $3$
(dot-dashed), $4$ (solid) as a function of
meson mode number $n$. }
}
\label{SpectrumFit}
\end{figure*}

\vskip -0.3cm
\begin{equation}
|<\Psi_{phy}|\Phi_g>|^2 = \frac{\alpha^2}{(M_- - m_2)^2}.
\end{equation}
The mixing probability is proportional to the
overlap  of these two wave functions (w.f.).
We calculate
 the  probability for no mixing, i.e., $P_{GM} =1- | \langle
\Psi_{phy}|\Phi_g\rangle |^2$.
As one can see in
 Fig.~\ref{SpectrumFit}, the
  the  mixing should occur when $n_g=2,3,4$ and the  meson mode numbers  
$n \sim 10,13,17$.
 This condition reduces the  overlap probability for
mixing dramatically. Therefore, we predict the existence of almost pure
glueball states, in the scalar sector, in the mass range above $2$ GeV.

\section{Conclusions}
\label{conc}

In this contribution we presented the applications of the GSW model
to  the glueball and
meson
spectra.
We saw that the proposed modification of the metric is
 fundamental to reproduce
experimental data with only two parameters.
We propose the inclusion of an additional free parameter dilaton to get binding 
potential for tachionic  
5-dimensional masses.
Excellent agreements with data are found.
For the pion, the SW dilaton has been properly modify to describe the
chiral symmetry breaking in the model. Also in this case the comparison with
data is quite good.
We
 conclude by remarking the capability of the model in reproducing
several masses of very different hadronic systems with only few universal  
parameters and therefore leading to a relevant predicting power.

\section{Acknowledgements}

This
work was supported, in part by the STRONG-2020 project of the European Unions
Horizon 2020 research and
innovation programme under grant agreement No 824093. The author thank the 
organizers of the `19th 
International Conference on Hadron Spectroscopy and Structure (HADRON2021)``.

\tabletopline\vspace{2pt}\lilahf{\sc Table I.\ {\rm   Comparison of the masses
of the ground states  for the odd 
spin
glueballs (in MeV). }}
\begin{center}
\small{\renewcommand{\arraystretch}{1.3}
\renewcommand{\tabcolsep}{1.35pc}

\begin{tabular}{| c | c | c | c |}
\hline
 $J^{PC}$ & My \cite{Meyer:2004gx} & Li \cite{LlanesEstrada:2005jf} &
 Our Work \cite{Rinaldi:2021dxh}
 \\ \hline
$1^{--}$ & $ 3240\pm 480$ &  395 &$3308 \pm 15$
 \\ \hline
$3^{--}$ & $ 4330\pm 460$ &  4150 &$4451 \pm 12$
 \\ \hline
$5^{--}$ &  &  5050 &$5752 \pm 10$
 \\ \hline
\end{tabular}
}
\end{center}

\tabletopline\vspace{2pt}\lilahf{\sc Table II.\ {\rm  Same of Table I for even
glueballs.  }}
\begin{center}
\small{\renewcommand{\arraystretch}{1.3}
\renewcommand{\tabcolsep}{1.35pc}
\begin{tabular}{| c | c | c | c |}
\hline
 $J^{PC}$ & My \cite{Meyer:2004gx} &Gy \cite{Gregory:2012hu} &Our Work
\cite{Rinaldi:2021dxh}
 \\ \hline
 $2^{++}$ & $2150 \pm 130$
& $2620 \pm 50$ & $2695 \pm 21$
 \\ \hline
 $4^{++}$ & $3640 \pm 150$
&  & $3920 \pm 14$
 \\ \hline
 $6^{++}$ & $4360 \pm 460$
&  & $5141 \pm 12$
 \\ \hline
\end{tabular}
}
\end{center}

\newpage

\vskip -0.5cm
{
\begin{table*}[h]
\tabletopline\vspace{2pt}\lilahf{\sc Table III.\ {\rm
Experimental results for
the $\pi$ masses given by the PDG particle
listings~\cite{Tanabashi:2018oca,Zyla:2020zbs} compared with our
calculations.
$\delta=1.5235$.  The masses  in MeV.  }}
\begin{tabular}{| c | c | c | c | c | c | c|}
\hline
& $\pi^0$ & & $\pi(1300)$ & & & $\pi(1800)$    \\ \hline
PDG & $ 134.9768\pm 0.0005$ &  & $1300\pm100$ &  & & $1819\pm 10$  \\ \hline
Our work \cite{Rinaldi:2021dxh} &   $ 135  $    &$943 \pm 111$&       $ 1231
\pm 133$& $  1463
\pm151$ & $1663 \pm168$ & $1842\pm183$\\ \hline
\end{tabular}
\end{table*}
}

\newpage

\end{multicols}
\medline

\bibliography{GSWPRD20213.bib}



\end{document}